\documentclass[twocolumn,superscriptaddress]{revtex4-1}

\usepackage{amsmath,amssymb}  
\usepackage{amsfonts} 
\usepackage{graphicx} 
\usepackage[dvipsnames,svgnames,x11names,hyperref]{xcolor}
\usepackage{siunitx}
\usepackage{subfig}
\usepackage{blindtext}
\usepackage[colorlinks=true, breaklinks=true, bookmarks=true, urlcolor=Blue, citecolor=OliveGreen, linkcolor=NavyBlue, bookmarksopen=false, draft=false]{hyperref}
\newcommand{\tarc}{\mbox{\large$\frown$}}
\newcommand{\arc}[1]{\stackrel{\tarc}{#1}}

\begin{document}


\title{Using smartphone photographs of the Moon to acquaint students \\ with non–Euclidean geometry}

\author{Hugo Caerols}
\email{hugo.caerols@uai.cl}
\affiliation{Facultad de Ingenier\'{\i}a y Ciencias, Universidad Adolfo Ib\'a\~nez, Santiago 7941169, Chile.}
\affiliation{Grupo de Observaci\'on Astron\'omica OAUAI, Universidad Adolfo Ib\'a\~nez, Santiago 7941169, Chile.}
\affiliation{IAU National Astronomy Education Coordinator (NAEC) Chile team.}

\author{Rodrigo A. Carrasco}
\email{rax@uai.cl}
\affiliation{Facultad de Ingenier\'{\i}a y Ciencias, Universidad Adolfo Ib\'a\~nez, Santiago 7941169, Chile.}
\affiliation{Data Observatory Foundation, Santiago 7941169, Chile.}

\author{Felipe A. Asenjo}
\email{felipe.asenjo@uai.cl}
\affiliation{Facultad de Ingenier\'{\i}a y Ciencias, Universidad Adolfo Ib\'a\~nez, Santiago 7941169, Chile.}

\date{\today}

\begin{abstract}
	Although they are sometimes considered problematic to grasp by students, the concepts behind non-Euclidean geometry can be taught using astronomical images. By using photographs of the Moon taken with a smartphone through a simple telescope, we were able to introduce these concepts to high-school students and college newcomers. 
	By recognizing different Moon geological structures within the photograph, we teach students how to calculate distances of mountain ranges or areas of craters on the Moon's surface, introducing the notions of geodesics and spherical triangles. Furthermore, students can empirically see that the correct estimations for the actual values cannot be obtained using flat geometry. Instead, by using three--dimensional curved geometry, precise estimates of lengths and areas of geological elements in the Moon can be computed with less than 4\% error. These procedures help students understand, concretely, non-Euclidean geometry concepts.
\end{abstract}
 
\maketitle 

\section{Introduction}
\label{sec:intro}
Although Galileo Galilei's {\em Sidereus Nuncius} \cite{galilei1610}, published in 1610, is one of the oldest scientific treatises about the Moon, humans have been fascinated and studied it intensively even before Euclid set the basis of Euclidean geometry in his famous {\em The Elements} \cite{euclid-300}. In a certain way, the Moon represents a reflection of our planet but, at the same time, something completely different. And the most noticeable feature of the Moon is that it is not flat.

Similarly, the Universe and its structures are not Euclidean, i.e., they are not flat in general. This idea is the cornerstone of the most influential theories in physics, including {\em General Relativity}, making the topic harder to explain to high-school and first-year college students \cite{Zahn_2014}. Our current understanding of the evolution of cosmology or astrophysical objects relies heavily on their description in curved spacetime, i.e., in their depiction in a non--Euclidean geometry in four dimensions. This fact is particularly true in astronomy, where the consideration of curved spaces and surfaces is essential \cite{smart1977textbook}.

The introduction of curved geometry to students is always a challenge; this difficulty is especially true for high-school students. For such topics, it is paramount to create engaging classes with clarity in their content \cite{Sirnoorkar_2016}. Our proposed introductory step to understanding the complexities of curved geometry is to make students deal first with the differences between three--dimensional Euclidean and non--Euclidean geometries. In many settings with challenging topics, the use of astronomy or astronomy-related activities has shown to significantly help high-school and college students grasp complex concepts and help them get into science \cite{hayes2020,rebull2018}. Furthermore, several researchers have shown that evidence-based active engagement courses help students significantly to grasp complex concepts \cite{lee1997,wells1995}, and even reduce the gender gap in STEM areas \cite{Karim2018}. This improvement is further enhanced by making students use real data and even capture the data themselves \cite{Boldea_2017,Farr2012}. Hence, what better element to introduce the differences between Euclidean and non-Euclidean geometries than a ``spherical" body, such as the Moon. 

In this work, we propose using low-resolution photographs of the Moon, taken with a smartphone and a simple telescope, to acquaint students with different concepts of three--dimensional curved geometry, such as distances and areas, and, at the same time, introduce concepts like geodesics. We use all these concepts to adequately explain how, using non--Euclidean geometry, we can measure different aspects of the Moon's topology. We show students how the approximations given by Euclidean geometry break in this setting, and how to extend those ``flat" concepts to a non-Euclidean setting, to improve the estimations significantly. Therefore, any student interested in how geological elements in the Moon can be measured will perform its own calculations with this proposal. 

We start in Section \ref {sec:photo}, showing that our natural satellite presents a valuable teaching tool to explain curved geometry. As it always shows the same face to Earth, due to tidal locking, any student can take photographs of it at any time to compare their results with known values. Next, in Section \ref {sec:flat}, we show through simple mathematical concepts, like {\em proportionality}, that estimations of distances in craters or valleys in the Moon are only valid for small segments or elements on its surface, compared to the Moon's radius.  Then, in Section \ref {sec:curveDistance}, we explain how the methods of non--Euclidean geometry are helpful when we are interested in estimating the length of long chains of mountains. In this section, we also show how the results differ when considering a flat geometry. In the last part of this work, we use those methods to calculate areas of different lunar maria by approximating such calculations by the sum of simple curved polygons. To successfully perform such a task, in Section \ref {sec:curveArea}, we introduce concepts of spherical trigonometry, allowing us to obtain any desired area as the sum of spherical triangles. These concepts have been used in the past to calculate the Moon's radius by counting its craters \cite{ardenghi2019estimation}, and we use them in Section \ref {sec:moonArea}, to estimate the areas of several different maria.

\section{Photograph of the Moon}
\label{sec:photo}

Our planet has the unique privilege in the Solar System of having a particularly large Moon, compared to our planet’s size. Furthermore, we always see the same face of the Moon due to tidal locking, allowing every viewer to see and photograph the same features.

\begin{figure}[b]
	\centering
	\includegraphics[width=8cm]{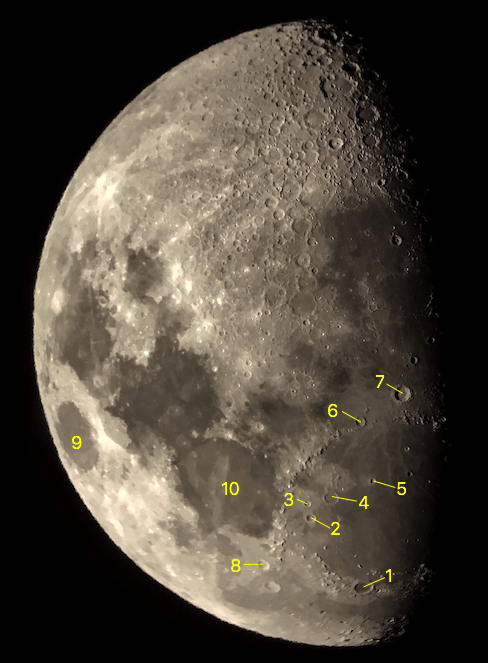}
	\caption{Moon right after first quarter,  9 days old and with a luminosity of 67.4 \%. Photograph taken with a smartphone and telescope on April 21, 2021 at 20:27, from Santiago, Chile. The numbers indicate craters and mare that we analyze in this study.}
	\label{figejemplo}
\end{figure}

In principle, any photographs of the Moon can be used. However, it is always significant that students participate in this activity, taking their own clear photographs of the Moon. The Moon experiences different phases during a month, translating into different amounts of luminosity that its surface reflects. To obtain the best photographs for these activities, in which information can be extracted, we recommend using photos taken when the Moon is not in a full moon phase. When in its full moon phase, the face of the Moon we see receives the light of the Sun straight into its surface, flattening its features and preventing the observation of specific details. We suggest planning the observation activity to take the photographs within a few days of  the first quarter phase, when the sun’s rays highlight the moon’s features. Additionally, the first quarter phase is visible in the hours just after sunset, which is a convenient time for most people and this side illumination highlights its features.

An example of a good photograph of the Moon is shown in Fig.~\ref{figejemplo}. The numbered craters in this figure are used in the following sections and they are described in detail in Fig.~\ref{figejemplo2}. It was taken using an iPhone 6, attached to the telescope with a NexYZ 3-Axis Universal Smartphone adapter. For this image, we used a Celestron Nexstar 8SE, which has a focal length is $2,032$ mm.  We use an ocular with a focal length of $25$ mm. We also used a standard {\em lunar filter} to cut down on the Moon's bright glare. Note that an even smaller and cheaper telescope, like a Celestron C90 with a focal length of $1,250$ mm, can also be used to obtain the desired images. After the photograph is taken, a smartphone app, such as Snapseed \cite{Snapseed}, can be used to enhance the contrast in the picture.

Although the orientation of the photograph is not essential for the proposed calculations, it should be noted that telescopes invert images and that they need to be mirrored or rotated to represent what is seen with the naked eye. Furthermore, the images of the Moon in the northern and southern hemispheres are reversed. This last point is relevant to consider since the image of the Moon in Fig.~\ref{figejemplo} was taken from the southern hemisphere.

\section{Small distances, flat geometry, and proportionality}
\label{sec:flat}

It is important to note that for small distances in the Moon's surface, compared to the Moon's radius, Euclidean geometry is enough to estimate the distance between two points accurately for points that are not near the limb. However, near the limb, Euclidean geometry fails, as we will see in the case of Plato crater (see below).

In this section, we will use simple Euclidean geometry to estimate distances. Through this process, we will also determine what we mean by ``small" distances. We will show in the following section how the Euclidean geometry approximation breaks down when elements with more considerable distances on the Moon's surface are considered.

\begin{figure}[t]
	\centering
	\includegraphics[width=7cm]{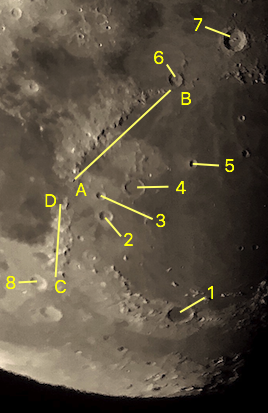}
	\caption{Zoomed in view of Fig.~\ref{figejemplo}. Eight craters with their respective names: (1) {\it Plato}, (2) {\it Aristillus}, (3) {\it Autolycus}, (4) {\it Archimedes}, (5) {\it Timocharis}, (6) {\it Eratosthenes}, (7) {\it Copernicus}, (8) {\it Eudoxus}.  The straight line between points $A$ and $B$ represent the approximated Euclidean flat distance of {\it Montes Apenninus}. The straight line between points $C$ and $D$ represent the approximated Euclidean flat distance of {\it Montes Caucasus}.}
	\label{figejemplo2}
\end{figure}

Consider Fig.~\ref{figejemplo2}, which shows a magnification of the lower part of Fig.~\ref{figejemplo}, where eight lunar craters are clearly distinguished. We can use proportionality to calculate the diameter of those eight craters. The real values of the crater's diameters $d_C$ are small compared to the diameter of the Moon $2 R_M$, where the Moon's radius is $R_M=1,737.4$ km \cite{moonFactSheet2021}. As a side note, although this parameter is well known, its value can be estimated in several simple manners, such as using smartphone photos of a lunar eclipse \cite{caerols2020estimating}. For cases such as these craters, the geometry of the crater can be considered Euclidean, and the proportions between the distances linear. Thus, a diameter of a crater can be computed as
\begin{equation}
	d_C= \frac{R_M }{R_{Mp}}d_{Cp}\, ,
	\label{calculatediameter}
\end{equation}
where $R_{Mp}$ is the Moon's radius in the photograph, and $d_{Cp}$ is the crater's diameter in the photograph. To obtain those values, we use GeoGebra \cite{GeoGebra2019} to analyze the photographs. With this software, we can calculate how large the diameter of the craters and the Moon's radius is the photograph on a centimeter scale. A detailed procedure to obtain those values can be found in 
\cite{caerols2020estimating}. 

Using the previous procedure to Figure \ref{figejemplo2} results in $R_{Mp}=5.3345$ cm. The same  can be performed to  obtain the diameters $d_{Cp}$ of different craters (listed in Table \ref{tabla1}).
Thus, through this procedure, any crater's diameter $d_C$ can be obtained by measuring its equivalent diameter distance in the photograph and multiplying the result by ${R_M }/{R_{Mp}} = 325.691255$ km/cm. In Table \ref{tabla1}, we show the seven different values $d_{Cp}$ of the crater's diameter obtained in the photograph (in cm), and the calculated diameters $d_C$ (in km) obtained from Eq.~\eqref{calculatediameter}. To measure the diameter of a crater in Geogebra, place two points that are diametrically opposite at the crater's edge. This choice should be made as precisely as possible, using the zoom in to correct those positions. This procedure helps reduce the errors in the measurements done in the software.

We  compare the estimated diameters with the exact values $d_{Cexact}$ obtained from the Working Group for Planetary System Nomenclature (WGPSN) from International Astronomical Union (IAU)\cite{Nomenclature2020}. The error between the actual value of the crater's diameter and the estimation done through our linear calculation is computed as $100|d_{Cexact}-d_C|/d_{Cexact}$. 

\begin{table}[t]
	\begin{ruledtabular}
		\begin{tabular}{l c c c c}
			Crater   & $d_{Cp}$ cm  & $d_C$  km & $d_{Cexact}$ km & error  \%\\
			\hline	
			(1) Plato & 0.3212 &  104.61  & 100.68  & 3.91\% \\
			(2) Aristillus &  0.1653 &  53.84 &54.37 &0.98\% \\
			(3) Autolycus& 0.1197& 38.99 &38.88 &0.27\% \\
			(4) Archimedes&  0.2476 & 80.64 & 81.04 & 0.49\%   \\
			(5) Timocharis &  0.1032 &  33.61 & 34.14 & 1.55\% \\
			(6) Eratosthenes & 0.1757& 57.22 & 58.77 & 2.63\% \\
			(7) Copernicus &  0.2893 & 94.22 & 96.07 & 1.92\%   \\
		\end{tabular}
	\end{ruledtabular}
	\centering
	\caption{Diameters for the seven craters of Fig.~\ref{figejemplo2}, estimated and real values.}
	\label{tabla1}
\end{table}

We will consider that a good estimation has an error of less than $1\%$, approximately. In Table \ref{tabla1}, we see
that the best results are obtained for the three smaller craters at the center of the photograph. 
On the other hand, the most significant error occurs for the Plato crater, with an error of $\sim4\%$. Note that this error can be partially attributed to its size and the lack of good luminosity at its position, close to the boundary of the moon picture. However, this error is already an indicator that the linear (i.e., Euclidean) approximation for these distances is already breaking down. For such cases, we need to consider the curvature of the Moon's surface, and thereby, use non--Euclidean geometry.

The size of the pixels in the photo used determines the uncertainty of the measurements. In this case, it measures $10.6 /2448=0.00433$ centimeters. When scaled to the size of the Moon, this value gives us between 1 and 2 km per pixel, depending on whether the crater is near the center or closer to the edge of the photo. Finally, we note that the errors obtained in Table I (and II) are in those ranges, which we consider an acceptable level for photos obtained with smartphones.

\section{Large distances, curved geometry and geodesics}
\label{sec:curveDistance}

For more considerable distances in a curved surface, using proportionality is insufficient to provide estimations with reasonable accuracy, as shown in the previous section. To calculate distances in a different geometry it is essential to know the concept of {\em geodesic}.

A geodesic is the line over the corresponding surface, between two points, with minimum length. This line can be in any geometry, flat or curved, and therefore, can be straight or curved. For example, given two points in Euclidean geometry, such as a plane, the geodesic is the straight-line between those two points. But for the curved surface of a sphere, the geodesic in this non--Euclidean space corresponds to the line that passes through the two points and, at the same time, is in a maximum circle. That is, the line is in the perimeter of a circle on the surface of the sphere.

For measuring large distances between two points in the Moon's surface is enough to approximate the Moon to a sphere and calculate the geodesics between such points. In what follows, we consider the actual Moon geographical values to be the ones appearing in Google Earth Pro free software \cite{GoogleEarth2020}.

As an example, let us first measure the {\it Montes Apenninus} length, the mountain chain shown in the photograph displayed in Fig.~\ref{figejemplo2}. The distance between points $A$ and $B$, considering a Euclidean space and the proportionality explained in the previous section, results in an estimation of 552 km approximately. However, the real value between those points is 600 km \cite{GoogleEarth2020}. This estimation has a $8\%$ error due to its geometry assumptions. This error occurs because we are not taking into account that Moon is spherically curved. 
Considering two points that are even further, the error grows significantly. For instance, the correct distance between craters (1) and (7) is 1,300 km \cite{GoogleEarth2020}. But using proportionality, we obtain an estimation of only 1,081 km. This estimation results in an error of about 17\%. 

To consider non-Euclidean geometry, we first note that  surface points are projected in a plane in the equator in any photograph of the Moon. Therefore, actual distances between two points in the photographs must be calculated by re-projecting them in the spherical surface of the Moon. Thus, the correct distance corresponds to the arc length of maximal circumference, i.e., the geodesic in the sphere.

\begin{figure}[h!]
	\centering
	\includegraphics[width=9cm]{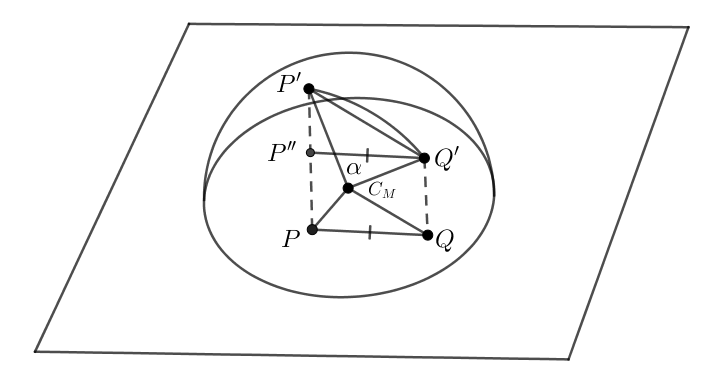}  
	\caption{Scheme where $P$ and $Q$ represent two points in the Moon photograph. Points $P'$ and $Q'$ correspond to their projections in the Moon's spherical surface. Other points are depicted in order to perform the calculation.}
	\label{figuratelescopesuncalproy}
\end{figure}

This is exemplified in Fig.~\ref{figuratelescopesuncalproy}.  In this Figure, $P$ and $Q$ represent two points in the photograph of the Moon. On the other hand, $P'$ and $Q'$ are the corresponding projection points in the Moon's surface, with  $C_M$ being the center of the Moon. The real distance will be the curved geodesic between points $P'$ and $Q'$. This  arc length can be calculated using Pythagorean theorem and trigonometry. This length is given by
\begin{equation}
	\overbrace{P'Q'}^{Geodesic}=\alpha R_M\, .
	\label{geode1}
\end{equation}
where $\alpha$ is the angle formed by the three points $P'$, $C_M$, and $Q'$,
\begin{equation}
	\alpha=\widehat{P'C_MQ'}=2\arcsin\left(\frac{d(P',Q')}{2R_M}\right)\, ,
\end{equation}
and $d(P',Q')$ is the  distance of the straight line between points $P'$ and $Q'$. Hereafter $d(A,B)$ is used to denoted the distance between any points $A$ and $B$. This distance can be obtained by the Pythagorean theorem as
\begin{eqnarray}
	d(P',Q')&=&\sqrt{d(P'',Q')^2+d(P'',P')^2}\, ,\nonumber\\
	&=&\sqrt{d(P,Q)^2+d(P'',P')^2}\, ,
\end{eqnarray}
as $d(P'',Q')=d(P,Q)$. Now, let us note that
\begin{equation}
	d(P'',P')=d(P,P')-d(Q,Q')\, ,
\end{equation}
and 
\begin{eqnarray}
	d(P,P')&=&\sqrt{R_M^2-d(C_M,P)^2}\, ,\nonumber\\
	d(Q,Q')&=&\sqrt{R_M^2-d(C_M,Q)^2}\, .
\end{eqnarray}

Using equation \eqref{geode1} we can obtain the real distance between two points over the Moon’s surface, by using the linear distance of those two points in the projection in the flat photograph, and their distances in the photograph to the center of the Moon, by using the following equation

\begin{widetext}
	\begin{equation}
		\overbrace{P'Q'}^{\textrm{geodesic}}= 2R_M \arcsin\left(\frac{1}{2R_M} \sqrt{d(P,Q)^2+\left( \sqrt{R_M^2-d(C_M,P)^2}-\sqrt{R_M^2-d(C_M,Q)^2}\right)^2}\right)\, .
		\label{geode2}
	\end{equation}
		\begin{table}[h]
	\begin{ruledtabular}
		\begin{tabular}{l c c c c c}
			Elements  & (A)  &  (B)   & (C) &  (D) & (E)\\
			\hline	
			Eudoxus,   point 8 in Figs.~\ref{figejemplo} and \ref{figejemplo2} & 47  & 68    & 67 & 1.49 & 30.9\\
			Montes Caucasus & 327 & 438   & 443 & 1.13 & 25.3\\
			Mare Crisium, point 9 in Fig.~\ref{figejemplo} & 203  & 546    & 555 & 1.62 & 62.8\\
			Montes Apenninus & 552 &  596     & 600 & 0.67 & 7.4 \\
			Distance between points 1 and 7 in Fig.~\ref{figejemplo2} &  1,081 &  1,302   & 1,300 & 0.15 & 17
		\end{tabular}
	\end{ruledtabular}
	\centering
	\caption{Different distances (in km) on the Moon's surface calculated using geodesic estimation. Column (A) are lineal (flat) distances, column (B) are curved distances \eqref{geode2}, column (C) are for Google Earth Pro (accepted) value, column (D) are the \% error between our estimation and accepted values, and column (E) are the \% error between linear and curved distance calculations. Mare Crisium  has an elliptical shape. We use the distance between points $V$ and $Q$ in figure \ref{figuratelescopesuncalmrecris}(a)  to calculate its radius, as explained later.}
	\label{bosons3a}
	\end{table}
\end{widetext}

This result can be easily programmed in GeoGebra to efficiently compute estimations for several structures on the Moon's surface. As an application, we can use equation \eqref{geode2} to estimate the real distances in Fig.~\ref{figejemplo2} for several points, and compare them with the values found in Google Earth Pro software \cite{GoogleEarth2020}.	
	
In Table \eqref{bosons3a} we summarize the values for several structures (in km). We show the estimation obtained considering a flat model, a curved model (computed with equation \eqref{geode2}), and the distance obtained in Google Earth Pro. We also calculate the error between both models (flat and curved) and the actual value obtained from \cite{GoogleEarth2020}. Notice how the flat approximation produces errors of 10\% or larger, whereas the curved one is significantly smaller.
This error is more significant if the geological elements are closer to the Moon's perimeter in the photograph, as is the case for the crater Eudoxus and Mare Crisium. As shown in Table \eqref{bosons3a}, the flat estimation is very poor when this occurs. Finally, note that there is uncertainty in these measurements due to the size of the pixel, which is between 1 and 2 km per pixel. This uncertainty is much smaller than the improvement achieved in the estimation when considering a curved surface over a flat one, which is the central objective of this activity.

\section{Calculating areas and the spherical triangle}
\label{sec:curveArea}

\begin{figure}[t]
	\centering
	\includegraphics[width=7cm]{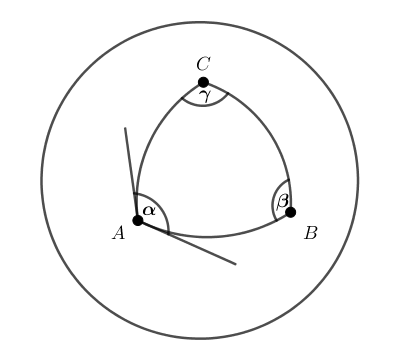}  
	\caption{Spherical triangle $ABC$ in a spherical surface, and its interior angles \cite{Brannan2011}}
	\label{figuratelescopesuncalaa}
\end{figure}

Now that students have been introduced to the concepts of curved surfaces and how to compute distances over them, we proceed to calculate a curved surface area using the flat projection of a Moon photograph. We apply this later to obtain the area of a polygonal region of the Moon, such as the 
Mare Crisium. We start with the basic spherical triangle in Figure \ref{figuratelescopesuncalaa}.

This figure shows three points in a sphere of radius $r$, together with the arcs of the maximum circles that pass through them. When the angles $\alpha$, $\beta$, and $\gamma$ are known, the area can be calculated straightforwardly using these angles. As shown in Ref.~\cite{Brannan2011}, the area of a curved triangle in the sphere is given by 
\begin{equation}
	a(\triangle ABC)=(\alpha+\beta+\gamma-\pi)r^2\, ,
	\label{aretrainguloesfericos}
\end{equation}
This results is known as Girard's Theorem. It also indicates that the sum of interior angles is always greater than $\pi$ in a sphere, something that does not occur in Euclidean geometry but is a staple of spherical geometry.

\begin{figure}[t]
	\centering
	\includegraphics[width=7cm]{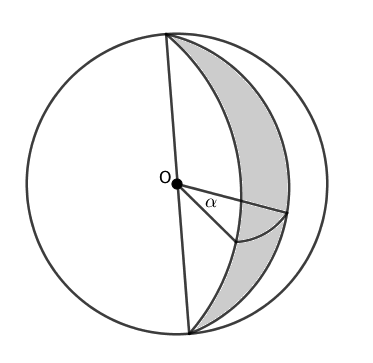}  
	\caption{Spherical spindle of angle $\alpha$ \cite{Brannan2011}.}
	\label{figuratelescopesuncalss}
\end{figure}

In order to demonstrate the equation for the area in \eqref{aretrainguloesfericos}, we first consider that the area of a spherical spindle is always proportional to its inner angle, that is $a(\alpha)=2\alpha r^2$ (see Fig.~\ref{figuratelescopesuncalss}).
When the angle is complete, i.e., $\alpha=2\pi$, then the area is the one of an entire sphere,  $4\pi r^2$. When $\alpha=\pi$, then we are dividing the sphere into two equal parts, with each area equal to $2\pi r^2$. We can continue in this way to realize that when we divide the total angle into $N$ equal parts, then the complete sphere splits into $N$ identical spindles and $N$ equal areas, completing the demonstration. 

\begin{figure}[t]
	\centering
	\includegraphics[width=8cm]{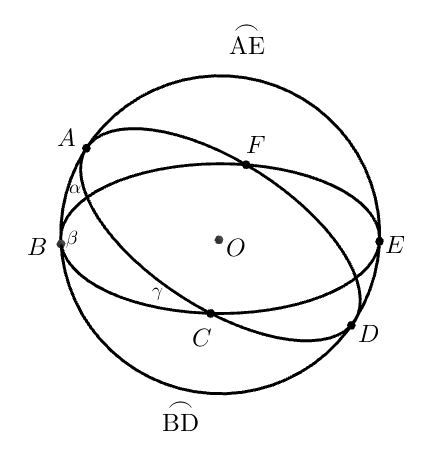}  
	\caption{Spherical triangles in a sphere \cite{calivieri}. Figure used to prove Girard's theorem.}
	\label{figuratelescopesuncal333}
\end{figure}

Let us now use this result to prove equation \eqref{aretrainguloesfericos}. To the best of our knowledge, this was first proved in \cite{calivieri}, as noted in \cite{Todhunter1886}. For this purpose, we will use Fig.~\ref{figuratelescopesuncal333}.
The three geodesics of maximum circles in the sphere define the spherical triangle with vertices $A$, $B$, and $C$. Note that ($BCD\arc{\mbox{BD}}$) denotes the area of the lower spherical polygon in the sphere, while ($ACE\arc{\mbox{AE}}$) denotes the area of the upper spherical polygon. Using the result for a spherical spindle, we find that
\begin{eqnarray}
	(ABC)+ (BCD \arc{\mbox{BD}}) &=&2\alpha r^2\, ,\nonumber\\
	(ABC)+ (ACE \arc{\mbox{AE}}) &=&2\beta r^2\, ,\nonumber\\
	(ABC)+ (ABF) &=&2\gamma r^2 \, .
\end{eqnarray}
However $(ABF)=(CDE)$ by symmetry (see Fig.~\ref{figuratelescopesuncal333}). Therefore, adding all the three previous equations, we obtain that 
\begin{eqnarray}\label{afsGSDG}
	&&3(ABC)+(BCD \arc{\mbox{BD}})+(ACE \arc{\mbox{AE}}) +(CDE)\nonumber\\
&&\qquad\qquad\qquad=2(\alpha+\beta+\gamma)r^2\, .
\end{eqnarray}
Summing together the three surfaces described in Eq.~\eqref{afsGSDG} comes out to be \begin{equation}2(ABC)+ \mbox{half of a sphere}=2(\alpha+\beta+\gamma)r^2\, .\end{equation}
Then, if we subtract half a sphere and divide by 2, we get the area  \eqref{aretrainguloesfericos}.

With this result, we only need to calculate the angles to calculate the area of interest. To do that, we can use the cosine rule for the sides of a spherical triangle \cite{Brannan2011}. This allows us to calculate the angles by knowing the distances between the vertices of a spherical triangle.

Let $ABC$ be the spherical triangle in Figure \ref{figuratelescopesuncalaa}, in which the sides ${AB}$, $BC$, and $CA$ have non-Euclidean lengths $c$, $a$, and $b$ respectively. The angles at $A$, $B$, and $C$ are $\alpha$, $\beta$, and $\gamma$ respectively. Then, we get that
\begin{eqnarray}
	\alpha &=&\arccos\left(\frac{\cos a-\cos b \cos c}{\sin b \sin c}\right)\nonumber\, , \\
	\beta &=&\arccos\left(\frac{\cos b-\cos a \cos c}{\sin a \sin c}\right)\, , \nonumber\\
	\gamma &=&\arccos\left(\frac{\cos c-\cos a \cos b}{\sin a \sin b}\right)\, .
	\label{anglescosinetheo}
\end{eqnarray}

The different lengths $a$, $b$, and $c$ can be calculated with the procedures developed in Sec.~\ref{sec:curveDistance}, and then, using Eqs.~\eqref{anglescosinetheo}, we obtain the angles. With this information, we can calculate the required area using Eq.~\eqref{aretrainguloesfericos}.

\section{Estimating  areas in the surface of the Moon using spherical triangles and polygons}
\label{sec:moonArea}

We are now in a position to compute an area on the Moon's surface using the methods from the previous sections. The previous formulas can be added to GeoGebra as code to compute the estimations in an easy way. We have left sample images and GeoGebra files in a public repository that can be found at 
\href{https://github.com/raxlab/moonGeometry}{https://github.com/raxlab/moonGeometry}
\cite{Caerols2021repository}.
Using this code, one can give three points in the photograph and, by using a spreadsheet, compute the total curved area of interest. 

\begin{figure}[b]
	\includegraphics[width=8cm]{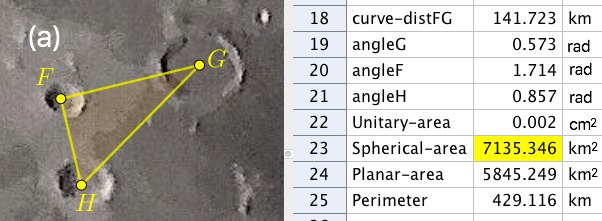} 
	\includegraphics[width=8cm]{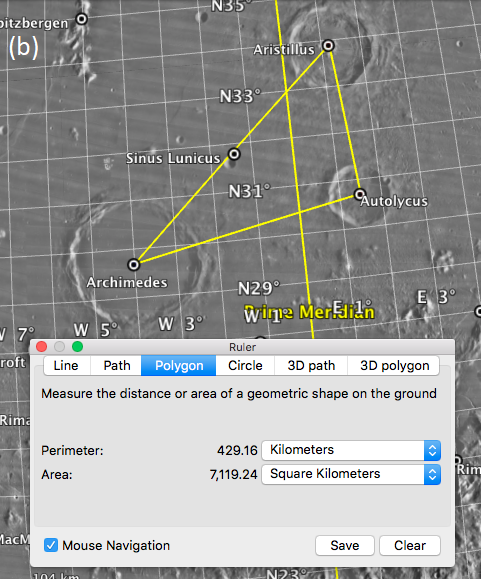}  
	\caption{(a) Zoomed in view part of photograph of Fig.~\ref{figejemplo2} showing the spherical triangle with vertices in the center of craters {\it Aristillus}, {\it Autolycus}, and {\it Archimedes}. Results obtained with the GeoGebra code are shown. (b) A spherical triangle is formed by the same craters constructed with Google Earth Pro software. The results for the curved areas using the software and our calculations are displayed in both figures for comparison.}
	\label{figuratelescopesuncalttt}
\end{figure}

Any total area can always be approximated as a sum of several spherical triangle areas. Therefore, the key is to estimate as accurately as possible the area of each spherical triangle. An example of the result is given in Fig.~\ref{figuratelescopesuncalttt}. 

A screenshot of the GeoGebra code is shown in Figure \ref{figuratelescopesuncalttt}(a). We calculate the area of the spherical triangle between the centers of the craters Aristillus, Autolycus, and Archimedes. These correspond to the points F, G, and H, respectively, which also correspond to the amplification of points 2, 3, and 4 in Figure \ref{figejemplo2}.  Part of the code used is depicted on the right-hand side of Figure \ref{figuratelescopesuncalttt}(a). We show the curved distance between points F and G, which is equal to 141.72 km, and the angles in each vertex, in radians. 

The normalized curved area is also shown, which, when multiplied by the Moon's radius (see Eq.~\eqref{aretrainguloesfericos}), results in the area for the spherical triangle, which results to be 7,135.35 km$^2$. As a comparison, the estimation using the flat area calculation only gives 5,845.25 km$^2$. To compare these estimations, we use the Google Earth Pro software to calculate the same spherical triangle among the same three points. This procedure is shown in Figure \ref{figuratelescopesuncalttt}(b) and results in a total area equal to 7,119.2 km$^2$. The estimation using curved areas obtained with our low-resolution photograph differs by only 0.2\% from the result using satellite photographs from Google Earth Pro. Both results show how different the measurements can be estimated when flat or curved geometries are used. 

Finally, the same two figures show how the curvature affects the distances. In Figure \ref{figuratelescopesuncalttt}(a), the perimeter of the triangle calculated by the sum of the three geodesics of each side (using Eq.~\eqref{geode2}), gives a result of 429.12 km. The same perimeter using Google Earth Pro results in 429.16 km.  These two results have practically no difference ($\sim 0.009$\% of difference).

The above comparison shows how good the calculation of curved geometrical structures is when using a low-resolution photograph taken by a smartphone. It is important to note that the procedure is sensitive to the position of the points. Thus, this must be taken into consideration when the distances and areas are calculated.

\begin{figure}[h!]
	\centering
	\includegraphics[width=5.7cm]{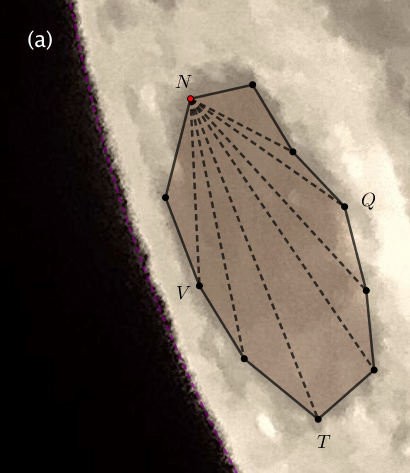}  
	\includegraphics[width=5.7cm]{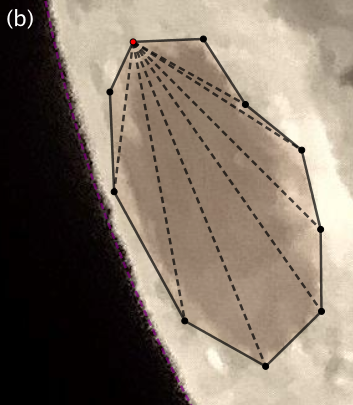}
	\includegraphics[width=5.7cm]{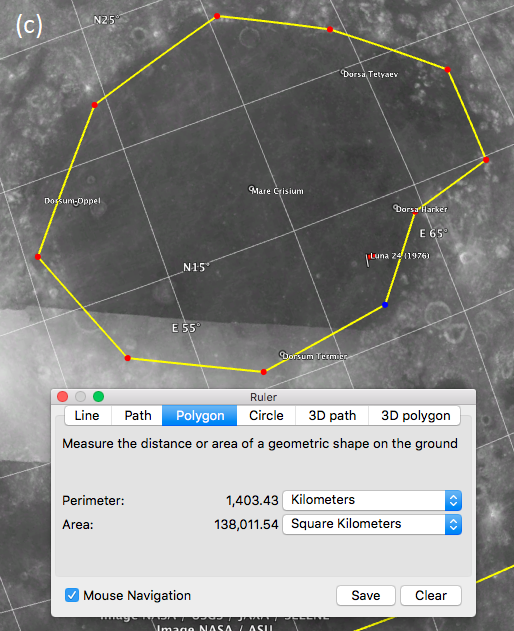}  
	\caption{Mare Crisium. (a) Interior triangulation to calculate its area, by using a ten side polygon. (b) Exterior triangulation with a  ten side polygon. (c)  Interior triangulation of the area, by using a ten side polygon, but with Google Earth Pro software.}
	\label{figuratelescopesuncalmrecris}
\end{figure}

We can use the above spherical triangle approach to obtain areas from the largest observable Moon structures, the Lunar Maria. We start by analyzing the Mare Crisium, which is shown as element 9 in Figure \ref{figejemplo}. In Figure \ref{figuratelescopesuncalmrecris}(a) we present an amplification of the sector that contains this mare.

The procedure is as follows. We can divide an area into several spherical triangles. The total area will be the sum of all the areas of the resulting triangle. This procedure is shown in Figure \ref{figuratelescopesuncalmrecris}(a). This image shows that we calculate the total area by estimating the mare's inner boundary with an interior triangulation, dividing it into ten triangles. Figure \ref{figuratelescopesuncalmrecris}(b) shows that we perform the same procedure but now for the outer boundary with an exterior triangulation. We do this to compute two bounds on the estimation for the total area of the mare. This procedure allows us to get a better approximation when using low-resolution photographs. Defining the correct boundaries in these types of photographs is very challenging. 

Using the previous procedure, we obtain a lower bound for the total area (with the inner boundary estimation) of 133,993 km$^2$. In comparison, the upper bound for the whole area (considering the outer boundary) is 205,684 km$^2$. Suppose we consider the estimation of the total area of {\it Mare Crisium} as the average of both bounds. In that case, we obtain an estimation of 169,839 km$^2$. This value is about 3.5\% off from the documented area of 176000 km$^2$ \cite{wilkinson2010moon}. Note that if we calculate these areas using a proportional approximation of the flat areas, we obtain as a lower bound of the area 48,497 km$^ 2$ and as an upper bound of the area 70,551 km$^2,$. Both are very far from the actual result.

Figure \ref{figuratelescopesuncalmrecris}(c) shows the same calculation performed with Google Earth Pro for the inner boundary, given a value of 138,012 km$^2$. Notice how close is this resulting value to our estimation for the inner boundary, being only 3\%. Furthermore, notice that in Figure \ref{figuratelescopesuncalmrecris}(a), the distance between points $V$ and $Q$ seems shorter than distance between $N$ and $T$. Actually, it is the opposite, as shown in Figure \ref{figuratelescopesuncalmrecris}(c). Our calculation using curved distances already computes that.

\begin{figure}[t]
	\centering
	\includegraphics[width=6cm]{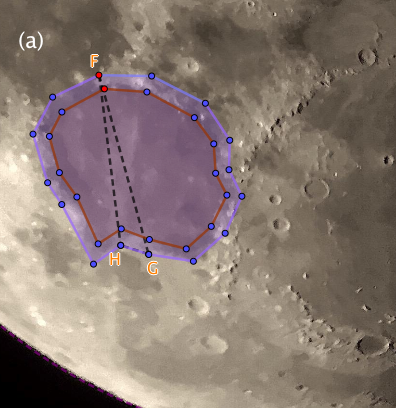}  
	\includegraphics[width=6cm]{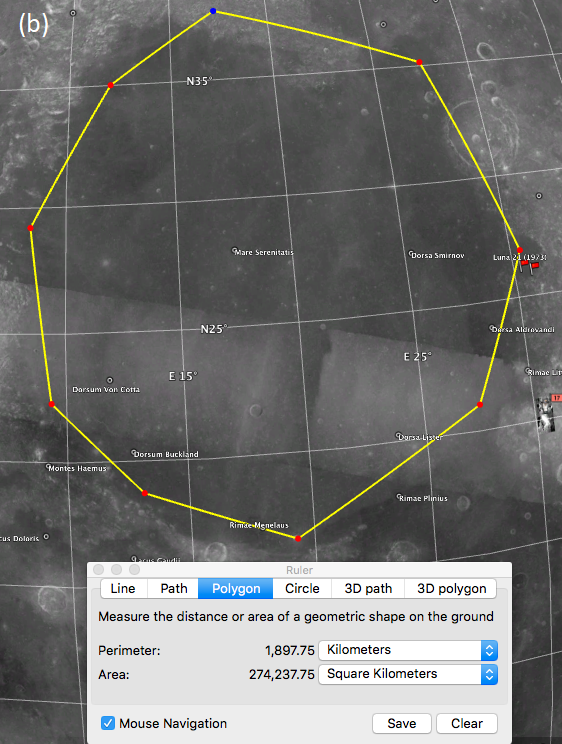}  
	\caption{Mare Serenitatis. (a) Interior and exterior triangulations to calculate its area, by using a fifteen side polygon. (b)  Interior triangulation of the interior area, but with Google Earth Pro software.}
	\label{figuratelescopesuncalsere}
\end{figure}

As a second example, let us focus on the {\it Mare Serenitatis} depicted as element 10 in Figure \ref{figejemplo}. Figure \ref{figuratelescopesuncalsere}(a) presents an amplification of the corresponding sector. In that image, we present the interior and exterior triangulation with fifteen triangles. For the inner boundary, we obtain an area of 261,959 km$^2$, whereas, for the outer boundary, we get an area of 457,504 km$^2$. The average of both bounds gives a total area of 359,731 km$^2$, which is only 1.8\% away from the real value of 353,400 km$^2$ \cite{whitford1982preliminary}. Figure \ref{figuratelescopesuncalsere}(b) presents the calculation for the inner boundary using Google Earth Pro, given an area of 274,237.75 km$^2$. Our result differs only a 4.5\% from that value.

In the above cases, for Mare Crisium and Mare Serenitatis, both the inner and outer boundaries result in very different estimations. For Mare Crisium, the difference in area between boundaries is 71,691 km$^2$ (corresponding to a difference of about 34\%), whereas, for the estimation of the area of Mare Serenitatis, the difference between boundaries is 195,545 km$^2$ (about 43\%). It is important to identify the origin of this uncertainty in the inner and out boundaries measurements. To see --and understand-- an image, we need approximately between $10^3$ \cite{Cai2003} and $10^4$ pixels \cite{Torralba2009}. With the same procedure used to measure the areas in  Fig.~\ref{figejemplo}, we can measure the equivalent area that corresponds to 1 pixel. We obtain that each pixel corresponds to roughly between 3 and 8 km$^2$. This difference implies that the minimum error that we can perform in the area estimation for our image can be of the order of $\sim 10^5$ km$^2$. Notice that this error for the area is of the same order that the ones obtained for both mares. Although the presented errors and uncertainties can be improved by choosing a better point to estimate the triangulations or increasing the resolution of the captured image, there is always an important error associated with the pixels of the final image.

\section{Conclusions}
\label{sec:conclusion}

This work shows how Moon photography can be used to teach non-Euclidean geometry. Although this subject can be harsh in general, and in particular for high-school students, we believe that by using their own photographs of the Moon, students can understand in a better manner the usefulness of curved geometry, and at the same time, realize that the components of nature (and Universe itself) are not generally flat.

The activity starts with a photograph of the Moon taken by the students themselves by using a smartphone. When distances and areas of some of the structures are calculated over the picture, for example, for the length of mountain chains or the size of craters, the student will realize that flat calculations differ enormously from the documented values in the literature. This insight is the key to introduce curved geometry, which can be developed using the same photographs and relatively simple mathematics.

To measure distances, the student will learn about geodesics on the surface of a sphere and how to use them. The basic concept is generally known from flat geometry. However, this time is extended to consider the curvature of the surface of interest. We further extend the study of non-flat surfaces calculate areas. For this, the concept of a spherical triangle is essential. 

The activity can also consist of creating a spreadsheet in Geogebra. This free software can automate the procedure using the points needed to construct distances and triangles. The students will also realize that their decisions when estimating lengths and areas will impact the final estimation, which is sensitive to the position of the selected points. This difficulty will put the delicate process to measure geological formations in other moons or planets into context.

As an initial complementary activity, the students can perform the same analysis for distances and areas for the photograph of a soccer ball or basketball. This exercise has the additional pedagogical purpose of comparing their calculations using the photograph and the above formulas with empirical measurements over the ball, which can be an engaging activity. Again, students will find that to get reasonable estimations; they have to consider the curvature of the surface. If they can program their spreadsheet in GeoGebra and retrieve their measurements for the ball, they will be able to do it with the Moon.

Furthermore, we have used parts of the approach described in this article in courses for high school math teachers, with excellent results that motivated teachers to adopt these strategies. We believe that the best way to use this approach is by challenging teams of learners. Depending on the context, the teacher can select either an active expository or a dynamic class. In the latter, teams of students are set to work consecutively on each of the topics we list in this article's different sections. According to the methodology used to address the problem, each covered topic can take between one or two 60-minute classes. For college-level courses, the teacher can do a guided activity as described above. In contrast, for summer workshops in astronomy, the teacher can do a research activity where the professor can exhibit the main questions or challenges of measuring distances and areas on spherical objects, and then challenge students to do so with a photo of the Moon, giving students the article as a reference to guide their work.

Finally, we believe that the current way to introduce the concept of curved geometry is, at the same time, challenging and motivating for the participants. Our experience has been precisely that for both students and teachers.

\begin{acknowledgments}
This research was partially funded by grants ASTRO020-0018, ASTRO20-0058, and FONDECYT 1180139 from ANID, Chile.
\end{acknowledgments}



\begin{thebibliography}{10}

\bibitem{ardenghi2019estimation}
Juan~Sebasti{\'{a}}n Ardenghi.
\newblock {An estimation of the Moon radius by counting craters: a
  generalization of Monte-Carlo calculation of $\pi$ to spherical geometry}.
\newblock {\em arXiv preprint arXiv:1907.13597}, 2019.

\bibitem{Boldea_2017}
Afrodita~L. Boldea and Ovidiu Vaduvescu.
\newblock {Teaching students about informatics and astronomy using real data
  for detection of asteroids}.
\newblock {\em European Journal of Physics}, 38(5):55706, aug 2017.

\bibitem{Brannan2011}
David~A. Brannan.
\newblock {\em {Geometry}}.
\newblock Cambridge University Press, 2nd editio edition, 2011.

\bibitem{caerols2020estimating}
Hugo Caerols and Felipe~A. Asenjo.
\newblock {Estimating the Moon-to-Earth Radius Ratio with a Smartphone, a
  Telescope, and an Eclipse}.
\newblock {\em The Physics Teacher}, 58(7):497--501, 2020.

\bibitem{Caerols2021repository}
Hugo Caerols, Rodrigo~A. Carrasco, and Felipe~A. Asenjo.
\newblock {\em {Moon Geometry Repository v.1.2}}.
\newblock https://github.com/raxlab/moonGeometry, 2021.

\bibitem{Cai2003}
Yang Cai.
\newblock {How Many Pixels Do We Need to See Things?}
\newblock In Sloot P.M.A., Abramson D., Bogdanov A.V., Gorbachev Y.E., Dongarra
  J.J., and Zomaya A.Y., editors, {\em Lecture Notes in Computer Science, ICCS
  2003}, pages 1064--1073. Springer Berlin Heidelberg, 2003.

\bibitem{calivieri}
Bonaventura~Cavalieri.
\newblock {\em {Directorium generale uranometricum}}.
\newblock 1632.

\bibitem{euclid-300}
Euclid.
\newblock {\em {The Elements}}.
\newblock 300 BC.

\bibitem{Farr2012}
Benjamin Farr, GionMatthias Schelbert, and Laura Trouille.
\newblock {Gravitational wave science in the high school classroom}.
\newblock {\em American Journal of Physics}, 80(10):898--904, 2012.

\bibitem{galilei1610}
Galileo Galilei.
\newblock {\em {Sidereus Nuncius}}.
\newblock 1610.

\bibitem{GeoGebra2019}
GeoGebra.
\newblock {GeoGebra Classic 5.0 User Manual}, 2019.

\bibitem{GoogleEarth2020}
Google.
\newblock {Google Earth Pro User Manual}, 2020.

\bibitem{hayes2020}
Christian~R. Hayes, Allison~M. Matthews, Yiqing Song, Sean~T., Linden, Robert~F.
  Wilson, Molly Finn, Xiaoshan Huang, Kelsey~E. Johnson, Anne~M. McAlister, Brian
  Prager, Richard Seifert, Sandra~E. Liss, Andrew~M. Burkhardt, and Nicholas
  Troup.
\newblock {First results from the Dark Skies, Bright Kids astronomy club
  draw-a-scientist test}.
\newblock {\em Physical Review Physics Education Research}, 16(1):10131, may
  2020.

\bibitem{Karim2018}
Nafis~I. Karim, Alexandru Maries, and Chandralekha Singh.
\newblock {Do evidence-based active-engagement courses reduce the gender gap in
  introductory physics?}
\newblock {\em European Journal of Physics}, 39(2), 2018.

\bibitem{lee1997}
Wei Lee, Heidi~L. Gilley, and Joshua~B. Caris.
\newblock {Finding the surface temperature of the Sun using a parked car}.
\newblock {\em American Journal of Physics}, 65(11):1105--1109, 1997.

\bibitem{moonFactSheet2021}
NASA Space Science Data~Coordinated Archive.
\newblock {\em {Moon Fact Sheet}}.
\newblock https://nssdc.gsfc.nasa.gov/planetary/factsheet/, 2021.

\bibitem{Nomenclature2020}
Gazetteer of~Planetary Nomenclature.
\newblock {International Astronomical Union (IAU) Working Group for Planetary
  System Nomenclature (WGPSN)}, 2020.

\bibitem{rebull2018}
Luisa~M. Rebull, Debbie~A. French, Wendi~Laurence, Tracyanne~Roberts, Michael~T. Fitzgerald, Varoujan~Gorjian, and Gordon~K. Squires.
\newblock {Major outcomes of an authentic astronomy research experience
  professional development program: An analysis of 8 years of data from a
  teacher research program}.
\newblock {\em Physical Review Physics Education Research}, 14(2):20102, jul
  2018.

\bibitem{Sirnoorkar_2016}
Amogh Sirnoorkar, Anwesh Mazumdar, and Arvind Kumar.
\newblock {Students' epistemic understanding of mathematical derivations in
  physics}.
\newblock {\em European Journal of Physics}, 38(1):15703, nov 2016.

\bibitem{smart1977textbook}
William~M. Smart and Robin~M. Green.
\newblock {\em {Textbook on spherical astronomy}}.
\newblock Cambridge University Press, 1977.

\bibitem{Snapseed}
Snapseed.
\newblock {Snapseed}, 2020.

\bibitem{Todhunter1886}
Isaac~Todhunter.
\newblock {\em {Spherical Trigonometry}}.
\newblock 1886.

\bibitem{Torralba2009}
Antonio Torralba.
\newblock {How many pixels make an image?}
\newblock {\em Visual Neuroscience}, 26(1):123--131, jan 2009.

\bibitem{wells1995}
Malcolm Wells, David Hestenes, and Gregg Swackhamer.
\newblock {A modeling method for high school physics instruction}.
\newblock {\em American Journal of Physics}, 63(7):606--619, 1995.

\bibitem{whitford1982preliminary}
James~L. Whitford-Stark.
\newblock {A preliminary analysis of lunar extra-mare basalts: Distribution,
  compositions, ages, volumes, and eruption styles}.
\newblock {\em The moon and the planets}, 26(3):323--338, 1982.

\bibitem{wilkinson2010moon}
John Wilkinson.
\newblock {\em {The Moon in Close-up: A Next Generation Astronomer's Guide}}.
\newblock Springer Science \& Business Media, 2010.

\bibitem{Zahn_2014}
Corvin Zahn and Ute Kraus.
\newblock {Sector models -- A toolkit for teaching general relativity: I.
  Curved spaces and spacetimes}.
\newblock {\em European Journal of Physics}, 35(5):55020, jul 2014.

\end{thebibliography}

\end{document}